\documentclass[reviewcopy]{elsart}

 \usepackage{graphicx}

\usepackage{amssymb}

\begin{document}

\begin{frontmatter}


 \title{Detecting long and short memory via spectral methods}
 \author{Simone Bianco}
 \ead{sbianco@unt.edu}
 \ead[url]{http://people.unt.edu/sb0269}
\thanks{The author would like to thank Prof. R. Ren\`o for research assistance and
 proofreading, and the Welch foundation for financial support through grant
 no. B-1577}




\address{Center for Nonlinear Science\\ University of North Texas, Denton,
  Tx, 76203-1427}

\begin{abstract}
We study the properties of memory of  a financial
time series adopting two different methods of analysis, the detrended
fluctuation analysis (DFA) and the analysis of the power spectrum (PSA). The
methods are applied on three time series: one of high-frequency returns, one
of shuffled returns and one of absolute values of returns. We prove that both
DFA and PSA give results in line with those obtained with standard
econometrics measures of correlation. 
\end{abstract}

\begin{keyword}
  serial correlation, high-frequency data, DFA, power spectrum, short memory, long memory.
\end{keyword}
\end{frontmatter}


\section{Introduction}\label{intro}
The analysis of financial time series has become an important field of application for
physicists. A special role plays the analysis of serial correlation in the
time series, as it gives useful insight to the price formation
mechanism. The study of serial correlation is not new, both in Econometrics
and Econophysics. The presence of serial correlation in the time series of
log-returns is forbidden by the efficient market hypothesis in its weak form,
see~\cite{fama}. At daily level the presence of these fluctuations has been
found in several markets and interpreted in the rational framework, with non-synchronous
trading~\cite{LoMac90} or institutional factors~\cite{BouRic94}, or invoking behavioral
factors~\cite{CutPot91}. \\ 
\indent The number of works on serial correlation with intraday and
high-frequency data is, at our knowledge, very limited due mainly to
the presence of microstructure effects in the time series that make difficult
a direct analysis of the problem. Recent examples are
Refs.~\cite{BiaRen06} on the Italian futures index,~\cite{LowMut96} on the
exchange rates and~\cite{ThoPat03} on the Indian market. In this paper we show how the adoption of popular
methods of statistical data analysis to infer about the presence of serial
correlations at intraday level, can be meaningful interpreted at the light of econometrics
measures, if the microstructure of the market is correctly taken into
account. Particularly, we adopt a simple analysis of the power spectrum  
(PSA) associated with the signal, and the famous Detrended Fluctuation Analysis (DFA)
method. \\
\indent The paper is organized as follows: Sections~\ref{PSA} and~\ref{DFA}
introduce the statistical instruments used, the data set is described
in Section~\ref{data}, Section~\ref{methodology}  shows the methodology that we 
employ and discuss the results at the light of
recent works on the same topic, while Section~\ref{conclusion} concludes.

\section{Power spectrum analysis}\label{PSA}
Let $x(t)$ be a stochastic process, we define the power spectrum as the
square modulus of the Fourier transform of the signal, namely:
\begin{equation}\label{pspectrum}
  G(f) = |\tilde{x}(f)|^2 .
\end{equation}
The power spectrum is linked to the autocorrelation function of the signal by
the Wiener-Khintchine theorem, namely by the following equation:
\begin{equation}
  \rho (\tau)  =  \int_0 ^\infty G(f) cos(2\pi f \tau) df
\end{equation}
which admits the following inverse relation:
\begin{equation}
  G(f) = 4 \int_0 ^\infty \rho (\tau) cos(2\pi f \tau) d\tau.
\end{equation}
In absence of correlations, the correlation function is $\delta-$peaked and
the power spectrum is flat (white noise). If on the contrary there is serial
correlation, this does not hold anymore and therefore we observe a
decay in the power spectrum  with the frequency as follows:
\begin{equation}
  G(f) \sim f^{-\eta}
\end{equation}
By monitoring the decay of $G(f)$ we can infer about the memory properties
of the time series under study. To perform the PSA we use a fast Fourier transform  algorithm.

\section{Detrended Fluctuation Analysis}\label{DFA}
The method of DFA has been widely adopted in the literature, starting from the
analysis of DNA sequences~\cite{dfa1} to financial time 
series~\cite{Matos}, from ecological applications~\cite{Telesca} to nuclear
reactions related problems~\cite{Alvarez}. It consists on the evaluation of the scaling properties of the locally
detrended standard deviation of the time series. We shall now briefly introduce
the algorithm.\\ 
Let again $x_i$, ($i = 1,\ldots, N$) be a stochastic process. The method consists on the
following steps:
\begin{itemize}
\item build the integrated time series $y_k = \sum_{i=1} ^N (x_i - \bar{x}), (k =
  1,\ldots,N)$ where $\bar{x}$ is the mean value of the signal;
\item divide the new sequence in $n = N/l$ non-overlapping subsequences of
  length $l$;
\item evaluate the local trend of each subsequence, $\bar{y}_k ^l(n)$;
\item evaluate the summation of the differences between the integrated time series and 
  the local trend in the time window $l$, and take its standard deviation,
  t. i.  the quantity:
\begin{equation}
  F(l) = \sqrt{\frac{1}{N} \sum_{k = 1} ^N [y_k - \bar{y}_k ^l(n)]^2}.
\end{equation}
\end{itemize}
The modified standard deviation $F(l)$ so built has the following scaling property:
\begin{equation}
  F(l) \sim l^{\alpha}.
\end{equation}
If $\alpha = 0.5$ the process is a white noise and there no
autocorrelation; if $0.5 < \alpha < 1$, then there is significant
autocorrelation in the time series. In the next Section we shall describe the
data set we use for the analysis of this paper.

\section{Data set description}\label{data}
The data set at our disposal consists of all the transactions taken from the
Italian futures on the stock index S\&PMIB, named FIB30, in the period from
January 2000 to December 2002. We use only the
next-to-expiration contracts. The data are evenly spaced according to a
previous interpolation procedure at a time lag of $1$ minute. The spacing has the effect of getting rid of
microstructure effects that can spoil the analysis, see Ref.~\cite{BiaRen06}
for a complete discussion. We have $751$ trading days, for a total of
$8,657,949$ transactions. After the spacing procedure, we have $495$
$1-$minute returns per day. The same data set has been also used in recent
publications, see Refs.~\cite{rosario, maria}.\\
In the next Section we shall see how the analysis
is performed.

\section{Methodology and results}\label{methodology}
\begin{figure}[!ht]
  \begin{center}
  \includegraphics*[height=6cm,width=12cm]{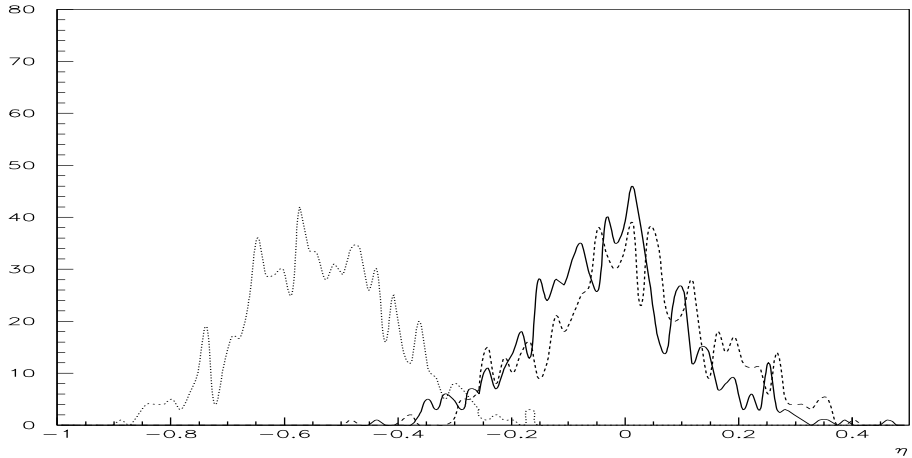}
\caption{In this Figure the distribution of the decay coefficients $\eta$
  obtained from the PSA algorithm on three time series: a time series of
  absolute returns (dotted line), a time series of shuffled returns (dashed
  line) and the original time series of returns (solid line).}\label{fft60}
\end{center}
\end{figure}

\begin{figure}[!ht]
  \begin{center}
  \includegraphics*[height=6cm,width=12cm]{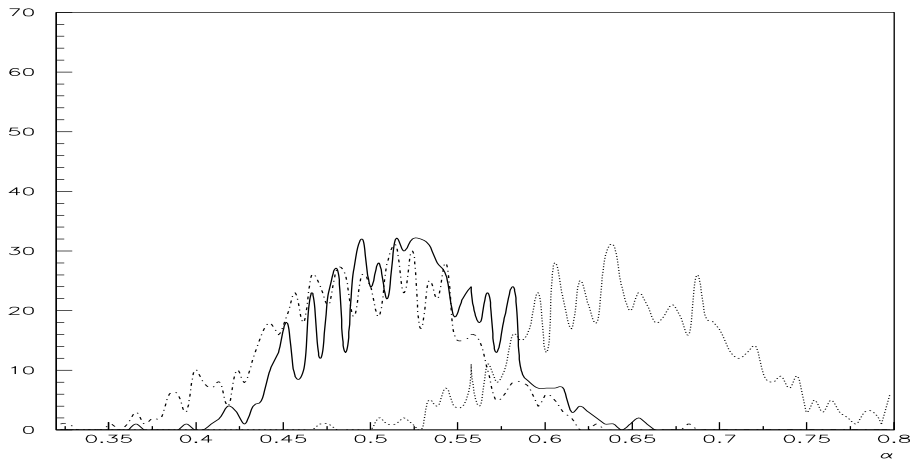}
\caption{In this Figure the distribution of the scaling coefficients $\alpha$
  obtained from the application of DFA on three time series: a time series of
  absolute returns (dotted line), a time series of shuffled returns (dashed
  line) and the original time series of returns (solid line).}\label{dfa60}
\end{center}
\end{figure}

In this Section we describe the methodology that we use to evaluate the
memory properties of the time series. We apply the methods introduced in the
previous Sections to three time series: one made of high-frequency returns,
whose memory properties are unknown, one obtained by shuffling the returns and
therefore without memory, and one of absolute returns, which are known to be
long range autocorrelated. We repeat the procedure for every day of our data set
and build the distribution of the coefficients obtained from the PSA and the
DFA. We then plot the three distributions and compare them to infer the
memory characteristics of the original time series. The results are
in Figs.~\ref{fft60} and~\ref{dfa60}. 

It is evident from the Figures that both the methods agree in detecting no
memory in the time series of shuffled returns: both distributions are centered
around the expected value, that is $0$ for PSA and $0.5$ for DFA. We note that the standard
deviation of the distributions is a reliable measure of the statistical
uncertainty of the coefficient. Moreover the results of the two methods when
applied to the time series of absolute returns are, as expected, compatible with
the hypothesis of long memory of this time series. When applied to the time
series of real returns, both DFA and PSA suggest the presence of short memory
effects in the time series, being the distributions of coefficients between
the other two. \\
In order to properly address the results we refer again to Ref.~\cite{BiaRen06}, where the
authors were able to prove, adopting  only econometric indicators, the presence of
short memory in the time series of returns. 
Our study therefore suggests that the
adoption of DFA and PSA is effective in the evaluation of short and long term memory in
financial time series.

\section{Concluding remarks}\label{conclusion}
In this paper we prove that an effective relation exists between what is
expected adopting scaling method of analysis, as DFA, spectral methods, as PSA, and
econometric indicators, as in Ref.~\cite{BiaRen06} in the evaluation of intraday serial correlations in
high-frequency financial time series. As far as our knowledge is
concern this study is the first to highlight this equivalence on
high-frequency data. We think that might be interesting to extend the analysis
on more liquid markets, as the US stock market, and we plan to address this
problem in the future.

\end{document}